\documentstyle[twocolumn,titlepage,openbib]{article}
\begin{document}
\newcommand{\N}{$^{1,2}$}
\newcommand{\m}{$\mu$m}
\renewcommand{\l}{$\lambda$}
\def\mincir{\raise -2.truept\hbox{\rlap{\hbox{$\sim$}}\raise5.truept
\hbox{$<$}\ }}
\def\magcir{\raise -2.truept\hbox{\rlap{\hbox{$\sim$}}\raise5.truept
\hbox{$>$}\ }}

\begin{titlepage}
\begin{center}
\vspace{1cm}
\Huge
{\bf
The Nuclear Ten Micron Emission of Spiral Galaxies
}

\normalsize

\vspace{4cm}

Giuliano Giuricin\N, Laura Tamburini\N,

Fabio Mardirossian\N, Marino Mezzetti\N, Pierluigi Monaco\N
\end{center}

\bigskip
\bigskip
\bigskip
\bigskip
\footnotesize

(1) Scuola Internazionale Superiore di Studi Avanzati (SISSA), via \linebreak
Beirut~4, 34013 -- Trieste, Italy

(2) Dipartimento di Astronomia, Universit\`a degli studi di Trieste, via
Tiepolo 11, 34131 -- Trieste, Italy

\normalsize
\bigskip
\bigskip

email: \\
giuricin@tsmi19.sissa.it\\
tamburini@tsmi19.sissa.it\\
mardirossian@tsmi19.sissa.it\\
mezzetti@tsmi19.sissa.it\\
monaco@tsmi19.sissa.it\\

\end{titlepage}

\begin{abstract}
\large

We examine the 10\m\ emission of the central regions of 281 spiral galaxies,
after having compiled all ground-based, small-aperture ($\sim$5") broad-band
photometric observations at \l$\sim$10\m\ (N magnitudes) published in the
literature. We evaluate the compactness of the $\sim$10\m\ emission of galaxy
nuclei by comparing these small-beam measures with the large-beam IRAS 12\m\
fluxes. In the analysis of different subsets of objects, we apply survival
analysis techniques in order to exploit the information contained in
``censored'' data (i.e., upper limits on the fluxes).

Seyferts are found to contain the most powerful nuclear sources of mid-infrared
emission, which in $\sim$1/3 of cases provide the bulk of the emission of the
entire galaxy; thus, mid-infrared emission in the outer disc regions is not
uncommon in Seyferts. The 10\m\ emission of Seyferts appears to be
unrelated to their X-ray emission.

HII region-like nuclei are stronger mid-infrared sources
than normal nuclei and LINER nuclei (whose level of emission is not
distinguishable from that of normal nuclei). Interacting objects have, on
average, greater 10\m\ luminosities than non-interacting ones and exhibit more
compact emission. Early-type spirals have stronger and more compact 10\m\
emission than late-type ones. Barred spirals are brighter at $\sim$10\m\ than
unbarred systems, essentially because they more frequently contain HII
region-like nuclei.

The results of our detailed comparison between the behaviour of various
categories of objects stress that the 10\m\ emission of spiral nuclei
is closely linked to the (predominantly non-thermal synchrotron) radio
emission.

\bigskip
\bigskip

{\it Subject headings:} galaxies: nuclei --- galaxies: Seyfert --- infrared:
galaxies

\end{abstract}

\normalsize
\section{Introduction}

Ground-based, small-aperture mid-infrared (MIR) photometric observations have
shown that the nuclei of spiral galaxies are often bright at these wavelengths.
Large observational efforts were performed particularly
by Rieke \& Lebofsky (1978), Lonsdale, Persson \& Mathews (1984),
Lawrence et al. (1985), Willner et al. (1985), Cutri \&
McAlary (1985), Devereux, Becklin \& Scoville (1987), Devereux (1987),
Ward et al. (1987), Carico et al. (1988), Wright et al. (1988),
Hill, Becklin \& Wynn-Williams (1988), Wynn-Williams \& Beclin (1993).
Strong MIR emission from galactic nuclei, which is clearly indicative of
non-stellar radiation, is frequently observed in Seyfert and starburst nuclei
(see, e.g., the review by Rieke \& Lebofsky, 1979), in several LINER nuclei
(Lawrence et al., 1985; Willner et al., 1985), in interacting and merging
galaxies (Lonsdale et al., 1984; Cutri \& McAlary, 1985; Joseph \&
Wright, 1985; Wright et al., 1988), with no obvious dependence on the spiral
morphological type (Devereux et al.,  1987), although it has been
claimed to be related to the presence of a bar, in early-type spirals
(Devereux, 1987). For a few galaxies mapped at fairly high resolution at MIR
and radio wavelengths, the
emission at \l$\sim$10\m\ shows a good overall spatial
correlation with the (thermal and non-thermal) radio emission (see, e.g., the
review by Telesco, 1988).

The MIR emission is generally attributed to thermal radiation from warm dust
which has been heated by early-type stars formed in a recent burst of star
formation, by shocks in supernova remnants and/or by a central dust-enshrouded
active galactic nucleus (AGN) (see, e.g., Ho et al., 1989 and the review by
Telesco, 1988); in the case of AGNs, the presence of a important non-thermal
power-law MIR emission component is conceivable, especially in the
high-luminosity objects, although in recent years its role has tended to
be deemphasized with respect to the thermal radiation
(see, e.g., the review by Bregman, 1990).

Spectroscopic observations in the MIR spectral band have revealed remarkable
differences in the main spectral properties of different categories of galactic
nuclei. The nuclei dominated by HII regions exhibit prominent narrow emission
features in the 3-13\m\ spectral interval, the so-called unidentified infrared
bands (UIR); these bands are generally absent in AGNs, which usually have
either featureless MIR spectra or spectra characterized by a strong silicate
[Babsorption band at \l$\sim$9.7\m. (Roche et al., 1991).

In view of the remarkable separation of the MIR spectra of different categories
of galactic nuclei, it is worthwhile to explore in detail how the strength of
the nuclear mid-infrared luminosity depends on the type of galactic nucleus and
on other characteristics of the host galaxy. This has not been
adequately examined in earlier MIR studies, although valuable efforts in this
direction have been already performed especially by Devereux et al. (1987),
Devereux (1987), and Hill et al. (1988).  Our investigation can help
us to cast more light on the nature of MIR emission in different classes of
objects.

In this paper, with the aim of providing a comprehensive picture of the MIR
activity in galactic nuclei, we examine the MIR emission of an extensive sample
of spiral galaxy nuclei. In Section \S 2 we present the adopted galaxy sample
and the small-aperture MIR (at $\sim$10\m) photometric data that we have
compiled from the literature. In Section \S 3, adopting a statistical approach,
we submit well-defined subsets of objects to a rigorous comparative analysis;
we use survival analysis techniques in order to exploit the information in
censored data (i.e. MIR fluxes). In Section \S 4 we discuss the main results of
our various two-sample comparisons and we address the relation between
10\m\ and X-ray emissions in Seyferts. Section \S 5 contains our conclusive
remarks.

\section{The Data Sample}

We have constructed our galaxy sample by choosing the spiral galaxies listed in
the ``Nearby Galaxies Catalog'' (NBG) by Tully (1988) with tabulated distance
$<$40 Mpc. This volume-limited catalogue is intended to include essentially all
known optically bright galaxies with systemic recession velocities of less than
3000 km/s. This corresponds to a distance of 40 Mpc for the Hubble constant
$H_0=75$ km s$^{-1}$ Mpc$^{-1}$, which value is adopted throughout the present
paper. In the NBG catalogue the distances of all non-cluster galaxies have been
essentially estimated on the basis of velocities, an assumed $H_0$ as above,
and
the Virgocentric retardation model described by Tully \& Shaya (1984), in which
the authors assume that the Milky Way is retarded by 300 km/s from the
universal Hubble flow by the mass of the Virgo cluster. The galaxy members of
clusters have been given a distance consistent with the mean velocity of the
cluster.

{}From the literature we have gathered together all published ground-based,
small-aperture, broad-band photometric measures at $\sim$10\m\ (N magnitudes)
for our galaxy sample. For the old literature we have consulted the reference
sources cited in the catalogue by de Vaucouleurs \& Longo (1988). In view of
the fairly large uncertainties associated with the N fluxes (which are
generally not smaller than 20\%) no corrections have been applied for the
redshift and the interstellar extinction. The central 10\m\ luminosity (in
solar units) is evaluated using the expression \begin{equation} L_N=8000 F_N
D^2 \end{equation} where $F_N$ is the 10\m\ flux measured in units of
millijansky and $D$ is the distance in units of Mpc (Scoville et al., 1983).

Table 1 lists the galaxies of our sample for which small-aperture $N$ data (or
upper limits) are available in the literature, together with the morphological
type T, bar type (SA, SAB, SB), and ring type S(r), S(rs), S(s) as coded in
the RC3 catalogue (de Vaucouleurs et al., 1991), the adopted distance $D$ (in
Mpc) as tabulated in NBG, the decimal logarithm of the central luminosity $L_N$
(in solar units) for detected objects or an upper (2$\sigma$) limit for
undetected galaxies, the projected linear size $A$ of the central region (in
kpc) corresponding to the beam size (mostly around 5"-6") used in the
observations, the reference source for photometry, the classification of the
nuclear emission-line spectrum as follows:H=HII region-like, L=LINER,
S1=Seyfert 1, S2=Seyfert 2, S=Seyfert; L/S (which indicates an uncertain
classification of the AGN spectrum); L/H (which indicates a composite AGN/HII
region-like nucleus in which the latter component is largely dominant over a
weak AGN); T (which indicates objects of transition between H and L). The
interacting galaxies are denoted by INT.
For the galaxies detected by the Infrared
Astronomical Satellite (IRAS) in the 12\m\ band, in the last column we list the
ratio (or its upper limit) of the ground-based small-beam 10\m\ flux to the
larger beam IRAS 12\m\ flux, colour corrected to 10\m. The resulting
compactness parameter $C$, which indicates the degree to which the central
region contributes to the MIR emission of the whole galaxy, is calculated
(following Devereux, 1987) as \begin{equation} C=(F_{10\mu m}/F_{12\mu m})
f_{cc} \end{equation} where the colour correction factor $f_{cc}$ is related to
the ratio between the IRAS 25\m\ and 12\m\ fluxes via the following relation
proposed by Devereux (1987): \begin{equation} f_{cc}=(0.12 F_{25\mu m}/F_{12\mu
m})+1.04\,. \end{equation} In a few cases in which we unreasonably obtain
$C>1$, we simply adopt $C=1$. Typical uncertainties in $C$ (due to errors on
fluxes) are on the order of 0.1.

We have included in our sample also a few galaxies with T=--1 (according to the
RC3 catalogue), which sometimes in the past have been classed as early-type
spirals. Priority is generally given to the most recent photometry and to the
measurements made with a beam size corresponding to a physical size of about
0.5--1 kpc (which is typical for most of the data).

For classification of the optical nuclear spectra we have generally relied
on the spectroscopic surveys of Stauffer (1982), Keel (1983, 1984), Keel et
al. (1985), V\'eron-Cetty \& V\'eron
(1986), Huchra \& Burg (1992), on the list of
starburst nuclei compiled by Balzano (1983), on the list of bona fide LINERs
culled by Willner et al. (1985), on the catalogues of Markarian objects by
Mazzarella \& Balzano (1986), on the listing of Durret (1989), and on the
catalogue of AGNs by V\'eron-Cetty \&
V\'eron (1991). According to the widely used
classification precepts of Baldwin, Phillips \& Terlevich (1981), the emission
line strength ratios H$\alpha$/[NII]\l 6584 and [OIII]\l 5007/H$\beta$ are used
in
order to distinguish Seyfert, LINER and HII region-like nuclei; for instance,
generally the HII region-like nuclei are characterized by line strength ratios
H$\alpha$/[NII]\l 6584 $>$ 0.7 and [SII]\l\l 6717,
6731/H$\alpha$ $<$ 0.4, whereas
LINERs have [NII]\l 6584 appreciably stronger than H$\alpha$ and [OIII]\l
5007/H$\beta$ $<$ 3. The unclassified nuclei are either objects with nuclear
spectra without emission lines or (mostly) objects with unobserved nuclear
spectra. A discrete fraction of the latter objects are likely to be LINERs in
the early-type spiral morphological range and HII nuclei in the late-type
interval, owing to the known preference of the two classes of nuclei to reside
in early- and late-type spirals, respectively (see, e.g., Keel, 1983).

We have denoted as interacting the galaxies listed in the ``Atlas and Catalogue
of Interacting Galaxies'' by Vorontsov-Velyaminov (1959, 1977), the interacting
galaxies culled by Davis \& Seaquist (1983) essentially from the UGC catalogue
(Nilson, 1973), the objects listed in the complete sample of interacting
galaxies surveyed by Keel et al. (1985) and in the sample of violently
interacting galaxies studied by Bushouse (1986, 1987), the paired galaxies
listed in the ``Catalogue of Isolated Pairs of Galaxies in the Northern
Hemisphere'' (Karachentsev, 1972, 1987), the binary galaxies selected by White
et al. (1983) and by Schweizer (1987), the twin galaxies identified by
Yamagata, Noguchi, Iye (1989). Our subsample of interacting galaxies is
likely to be skewed towards pairs of galaxies which are close to one another,
independently of their morphological appearance (several galaxy pairs show
no evidence of tidal distortion or structural peculiarity). Disturbed
galaxies with no apparent companions tend to be unrecognized as interacting
(see, e.g., the case of Maffei 2 discussed by Hurt et al., 1993). It is not
easy to understand whether this bias (which is common to most statistical
studies on the effects of galaxy interactions) can affect significantly
the results, since we would need to have a (presently unavailable)
wide sample of interacting galaxies
which are selected solely on the basis of the presence of morphological
features unmistakeably associated with a certain range of tidal strengths.

We have taken the IRAS 25\m\ and 12\m\ fluxes directly (in order of preference)
from the papers of Soifer et al. (1989), Helou et al. (1988), Rice et al.
(1988), Devereux (1987), which generally report global coadded fluxes from all
IRAS survey data, and from the ``Cataloged Galaxies and Quasars Observed in the
IRAS survey'' (Fullmer \& Lonsdale, 1989), which generally reports point source
fluxes; for NGC 4151 we have used the IRAS pointed observations of Edelson \&
Malkan (1987).

Our sample, which is the most extensive --- albeit not complete --- set of
small-beam MIR photometric data so far examined in the literature, can be
regarded as representative of the 10\m\ luminosity of the central regions
(of $\sim$0.5-1 kpc, typically) of nearby spirals. The inhomogeneity of our
data set is limited by the fact that about half of it comes directly from the
photometric survey of Devereux (1987) alone. We purposely avoid choosing
galaxies by virtue of some unusual property (e.g, infrared-bright galaxies,
peculiar and interacting galaxies, galaxies with active or starburst nuclei).
Therefore, our galaxy sample is less biased towards infrared bright galaxies
and, in general, towards exceptionally bright (in the optical and infrared
spectral bands) objects than several samples used in previous MIR studies (cf.
the infrared-selected galaxy samples of Devereux (1987), Carico et al. (1988),
Wright et al. (1988)). Yet our sample is likely to be still a little biased
in that sense (compared to a hypothetical complete sample), simply because
the attention of many observers was focused on those objects. This bias
implies that Seyferts (8\% ) and interacting objects (29\%) are
over-represented in our sample; it also probably leads to a general
underestimation of the number of nuclei of low 10 \m\ luminosity, although
it should not affect relative comparisons between the various
categories of objects.

\section{Statistical Analysis}

Owing to the fairly large number of upper limits on the MIR flux densities, we
have made use of statistical techniques suitable for the analysis of censored
data and adapted to astronomical usage from the field of survival analysis
(see, e.g., Schmitt, 1985; Feigelson \& Nelson, 1985; Isobe, Feigelson, Nelson,
1986; for extensive discussions on the astronomical applications). If the
non-detected galaxies were dropped from the sample (as was done in previous
efforts to characterize the typical MIR luminosities
of galactic nuclei), the resulting
sample would be skewed towards infrared-bright galaxies in a complicated
fashion that depends on the MIR luminosity function and on the sensitivity
limits. In practice, we have used the software package ASURV (Rev 1.0) (Isobe
\& Feigelson, 1990) which was kindly provided to us by E. Feigelson.

We have employed the Kaplan-Meier product-limit estimator in order to calculate
the means, medians, and cumulative distribution functions of the quantity log
$L_N$ for various subsets of galaxies. The Kaplan-Meier estimator, which is
the cornerstone of the non-parametric survival analysis, had been proved to
be the unique self-consistent, generalized maximum-likelihood,
asymptotically normal estimator (under quite broad conditions) of the
empirical distribution functions of a randomly censored data set.
The estimated cumulative distribution function (though not the associated
errors) is identical to the iterative solution derived by Avni et al.
(1980); it is a piecewise continuous and piecewise constant function
with jumps at the uncensored (detected) values. The Kaplan-Meier estimator
adequately treats detections and redistributes upper limits, recovering
the information lost by censoring; it has been shown
to be valid under quite diverse conditions (Feigelson \&Nelson, 1985).
The median of the distribution function is always well-defined. If the
lowest point in the sample is an upper limit, the mean is not well-defined,
since the distribution is not normalizable, and so the the outlying
censored point is redefined as a detection in our analysis.

The comparison of the distribution functions of log $L_N$ of
pairs of subsamples is accomplished by using two versions of Gehan's test (one
with permutation variance and the other with hypergeometric variance),
the logrank, the Peto-Peto, and the Peto-Prentice tests.
These tests differ in how the censored points
are weighted and consequently have different sensitivities and efficiencies
with different distributions and censoring patterns.

The 10\m\ emission of many galaxies is characterized by an appreciable spatial
extent ($>$ 0.5 kpc), as was stressed by the high-resolution
10\m\ maps of Ho et al. (1989) and by the small-beam 10\m\  photometry of
Devereux (1987), Wright et al. (1988), Carico et al. (1988),  Hill, Becklin \&
Wynn-Williams (1988), which is frequently compared with the large-beam IRAS
12\m\ fluxes. Thus, log $L_N$ is expected to increase as the beam diameter $A$
(i[Bn kpc) grows. This is illustrated by the log $L_N$ -- log $A$ plot (Fig.
1).
Fig. 1 shows that log $L_N$ increases approximately linearly with log $A$ up to
$A\sim 1$ kpc (which can be taken as a typical lower limit for the spatial
extension of the 10\m\ emission). In order to calculate the linear regression
line in this case, in which upper limits on the independent variable are
present, we have used three methods: the EM algorithm,
which requires that the functional form of the distribution
of the dependent variable about the regression line be
specified (a normal distribution is generally assumed), the non-parametric EM
algorithm with Kaplan-Meier estimator, suggested by Buckley \& James (1979),
and
Schmitt's method (see, e.g., Isobe et al., 1986). All three methods concur
that the slope of the log $L_N$ -- log $A$ relation is 1.7$\pm$0.2, i.e. close
to 2 (the value typical for a constant surface brightness source).  More
specifically, these considerations hold for the non-Seyfert objects, since the
23 Seyferts exhibit no significant log $L_N$ -- log $A$ correlation. This would
suggest that the MIR emission of Seyferts is not so extended as that of
the other galaxies. This issue will be better addressed below in the discussion
of the compactness parameter $C$. In the comparison of the log $L_N$
distributions of pairs of galaxy subsamples, one should take into account the
bias due to the log $L_N$ -- log $A$ relation by comparing subsamples having a
similar range of $A$-values. In practice, we deem it enough to exclude
objects with $A<0.3$ kpc in order to minimize this bias.

As expected, the compactness parameter $C$ shows a tendency to
increase with growing $A$. The $C$ -- $A$ plot is shown in Fig. 2; in
this figure the dashed lines represent a grid of simple galaxy
models in which the surface brightness decreases as $e^{-r/rs}$, where
$r_s$ is the characteristic radius of the surface brightness distribution.
This plot is similar to that introduced by Hill et al. (1988), except
that in this case we are plotting aperture diameters instead of distances.
Seyfert galaxies are characterized by smaller scale sizes ($r_s$) than
the norm; most objects cluster around scale sizes $\magcir 0.5$ kpc. Also
in the analysis of the $C$-distributions, we shall exclude the objects with
$A<0.3$kpc.

We define the MIR central surface brightness as $S_N=L_N/A^2$ (where $S_N$ is
expressed in units of solar luminosity per square kpc). The quantity log $S_N$
exhibits no significant correlation with $A$ (see the log $S_N$ -- log $A$ plot
illustrated in Fig.3), which is consistent with the fact that the 10\m\
emission is generally extended. We have verified the absence of correlation by
computing the correlation probabilities based on the Cox proportional hazard
model, on the generalized Kendall rank correlation statistics, and on the
generalized  Spearman rank order correlation coefficient. For $N_T=281$ objects
with $N_L=137$ upper limits on log $S_N$ we have found high probabilities that
the Btwo variables log $S_N$ and log $A$ are independent (p=0.47, 0.89, 0.79,
for the three respective tests). Therefore, in comparing the log $S_N$
distributions of pairs of galaxy samples we shall consider all objects
irrespective of their $A$-values.

The compactness parameter $C$ shows a moderate correlation with log $L_N$ and
log $S_N$ (see Figs. 4 and 5). This is essentially due to the fact that
the brightest categories of nuclei (at MIR wavelengths) -- such as Seyferts,
HII nuclei, and interacting nuclei -- tend to have also more compact emission
than the norm. (This is discussed in detail in the following section.)

For various subsamples Table 2 contains the main results of our statistical
analysis: the total number $N_T$ of (detected and undetected) objects, the
number $N_L$ of upper limits (undetected galaxies), the estimated medians,
means, and standard deviations $\sigma$
of the distribution functions of log $L_N$
(for $A\geq0.3$ kpc ), log $S_N$ (for all values of $A$), $C$ (for
$A\geq0.3$ kpc). Table 3  gives the numerical outcomes from the comparison of
the distributions of many pairs of galaxy subsamples, namely the (two-tailed)
probabilities p that two data subsets come from the same underlying population.
In Table 3 we give the mean ( $p(mean)$), minimum ($p(min)$), and maximum
($p(max)$) values of the five probabilities coming out from the application
of the five above-mentioned tests. The statistical
significance of the difference between the distributions of two
data subsets is at the 100$\cdot$(1-p) per cent level. These five tests
give somewhat different results especially when the data set is small
or suffers from heavy censoring.

The main results which can be inferred from an inspection of Tables 2 and 3 are
discussed below. In all Tables $L_N$ is expressed in units of solar luminosity
and $S_N$ in units of solar luminosity per square kpc.

\section{Results and Discussion}

\subsection{Seyferts and non-Seyferts}

First of all, it is very clear that the Seyfert nuclei (denoted by S in the
Tables) are unequivocally the most powerful MIR sources, since they have
typically greater 10\m\ luminosities and surface brightnesses than the
non-Seyfert ones (NON-S) (from which we have excluded the objects classed as
L/S in order to be conservative). Seyferts are also intrinsically brighter at
10\m\ than the two individual classes of HII region-like (H) and LINER (L)
nuclei. Our new finding, which indicates an excess MIR emission of Seyferts
with
respect to non-Seyferts, appears to be consistent with the results of
Spinoglio
and Malkan's (1989) study of an extensive 12\m\ flux-limited sample of IRAS
sources. Emphasizing the use of the IRAS 12\m\ flux as a good index of
non-stellar emission and as a powerful technique of AGN selection, the two
authors found that Seyfert galaxies have, on average, higher 12\m\
luminosities than non-Seyferts. Our analysis provides  decisive proof of
the predominance of Seyferts in the MIR emission, since IRAS beams, being about
a few arcmin in size, encompass a substantial fraction of the entire galaxy
(except in very nearby objects) and are thus less appropriate to the study
of galactic nuclear emission.

Seyferts are characterized by a much more compact MIR emission than the
other categories of objects, although in many (15) Seyferts (68\%) the
compactness parameter $C$ is low enough ($C<0.7$) to indicate the presence of a
substantial extended emission. This result is not appreciably affected by the
possible presence of a 9.7\m\ absorption silicate feature, which would
suppress the ground-based 10\m\ flux proportionally more than the IRAS 12\m\
flux (because of the wider bandpass of the IRAS 12\m\ filter). Hill et al.
(1988) have evaluated the amount of the consequent reduction of the compactness
parameter $C$ for a different spectral index of the power-law continuum of the
source and different optical depth at 9.7\m. According to their calculations,
for a quite negative spectral index (i.e. $F_\nu\propto\nu^n$ with n=--2), the
effect is a 13\% and 24\% decrease of C for $\tau_{9.7\mu m}$=1 and 2,
respectively (greater spectral indices lead to smaller effects). The
observations of the MIR spectra of galactic nuclei show that most of the
galaxies having this silicate absorption band,
which is thought to arise in dusty regions fairly
close to the central source, are type 2 Seyferts (Roche et al., 1991), for
which $\tau_{9.7\mu m}<2$ is a typical value (e.g., Roche et al., 1984). Even
if
we hypothesize an underestimation of the parameter $C$ by 25\% for all the
Seyferts of our sample which are not classed as Seyferts 1, we still find 13
objects (59\%) with fairly low $C$ ($C<0.7$). This indicates that extended
thermal MIR emission from dust warmed by hot stars in the outer regions (beyond
$\sim$0.5 kpc from the center) is important in the majority of Seyferts, in
agreement with the abundance of HII regions (in the galactic disks) which are
observed in optical emission-line surveys (e.g., Pogge, 1989).

Interestingly, both the subset of far-infrared luminous galaxies (of
the IRAS Bright Galaxy sample) observed by Carico et al. (1988) in the
near- and mid-infrared and the sample of 19 IRAS luminous galaxies,
observed by Wynn-Williams \& Becklin (1993) in several MIR bands, exhibit, on
average, somewhat more compact 10\m\ emission than our sample of Seyferts.
According to Carico et al. (1988), who compared their own small-beam 10\m\
measures with the IRAS 12\m\ fluxes, half of these galaxies have
10\m\ emission consistent with a contribution of 50\% or more from a
central point source. Wynn-Williams \& Beckin (1993), who measured
directly the compactness at \l$\sim$12\m\ and \l$\sim$25\m\ through
observations made with filters centered just at these two wavelengths,
foundB $C\geq 0.5$ for all galaxies (at both wavelengths).
The great compactness of the MIR emission of these infrared bright
galaxies may be related to their particularly great infrared
luminosities, with which the degree of compactness is found to correlate
(Carico et al., 1988).

The great predominance of Seyferts in MIR emission is consistent with
their well-known strong predominance in near-infrared emission (e.g.,
Glass \& Moorwood, 1985), in radio continuum emission (e.g., Ulvestad,
Wilson \& Sramek, 1981; Giuricin, Mardirossian \& Mezzetti, 1988b), and
in soft X-ray emission (e.g., Kriss, Canizares \& Ricker, 1980, and
Giuricin et al., 1991 for Seyferts 1 and 2 respectively). Furthermore,
the IRAS data indicate that the Markarian Seyferts exhibit also an excess
of 25\m\ flux with respect to the Markarian starbursts galaxies (Hunt,
1991). On the contrary, the far-infrared luminosities of these two
classes of objects are similar (Rodriguez-Espinosa, Rudy \& Jones, 1987;
Dahari \& De Robertis, 1988).

\subsection{Interacting and non-interacting galaxies}

Among the non-Seyferts (from which we have excluded the objects classed as L/S)
the interacting objects (denoted by INT in the Tables) exhibit greater values
of $L_N$ and $S_N$ in general than the non-interacting (NON-INT) galaxies, in
qualitative agreement with earlier results (Lonsdale et al., 1984; Cutri \&
McAlary, 1985; Wright et al., 1988). However, these studies tend to be biased
towards the most extreme cases of strong interactions and do not properly
include the upper limits in their analysis; therefore, they tend to
overestimate
the $L_N$ differences between the interacting and non-interacting galaxies. For
instance, Wright et al. (1988) reported a ratio of $>$10 between the $L_N$
averages of the two categories, whilst our Table 2 gives a ratio of $\sim$2
only ($\sim$1.7 for the $S_N$ averages).

In addition, we are able to explore the effects of interactions on the
intensity of the MIR emission separately for the two classes of LINER and HII
region-like nuclei. Tables 2 and 3 reveal that the difference between the
interacting and non-interacting LINERs (denoted by INT (L) and NON-INT (L)
respectively) is considerable in the case of the $S_N$ distributions (it is
small in the case of the $L_N$ distributions for $A\geq0.3$ kpc, because of
poor
statistics). On the other side, the interacting and non-interacting HII nuclei
(denoted by INT (H) and NON-INT (H)) differ very little in the $L_N$
distributions and not at all in the $S_N$. We have verified that the reason
for this behaviour of the HII nuclei stems from the fact that the $L_N$-values
of the interacting HII nuclei generally refer to greater $A$-values than the
non-interacting do. (The median values of A are 0.71 kpc and 0.60 kpc,
respectively.) On the reasonable hypothesis that a
discrete number of the unclassified nuclei hosted
in late-type spirals are likely to be HII nuclei, we have tried to improve the
statistics by adding to our sample of HII nuclei all unclassified galaxies of
type
Sc and later. This enlarged data sample confirms the marginal difference
between the $L_N$ distributions of the interacting and non-interacting objects
(denoted by INT (H+LATE) and NON-INT (H+LATE) respectively in Tables 2 and 3)
and the absence of difference in the $S_N$ distributions. Hence, some
interaction-induced enhancement of the MIR emission in HII nuclei may be caused
by a spread of the MIR emission over a somewhat larger area (which may be due
to an enlargement of starburst regions) rather than to an increase in the
central MIR surface brightness, as occurs in the case of interacting LINERs.
Notably, the difference in the interaction effects on the MIR emission of
the two categories of nuclei bears considerable similarity with the
interaction-induced enhancement of the central radio emission of LINER
and HII nuclei (see Giuricin et al., 1990).

Interacting objects also exhibit more compact MIR emission than
non-interacting ones. This holds for both HII and LINER nuclei. This finding
specifies that the interaction-induced enhancement of the MIR emission concerns
preferentially the central galactic regions rather than the outer disc regions
(as first suspected by Lonsdale et al. (1984)), although the  compactness
parameter $C$ is great ($C>0.5$) only in eight (18\%) interacting objects.

The increased compactness of the 10\m\ emission of interacting objects
agrees with the similar behaviour of various indices of star formation
activity: H$\alpha$ line emission (Bushouse, 1987), radio emission
(Hummel, 1981b; Condon et al., 1982), far-infrared emission (Bushouse,
1987), and near-infrared emission (Joseph et al., 1984; Cutri \& McAlary,
1985); in the latter case the effect was found to be detectable in HII
nuclei only (Giuricin et al., 1993). All these results are generally
taken as an indication that current star formation activity (induced
by interactions) preferentially occurs in and near the nuclear regions
of galaxies rather than in their outer disk regions (see also
Laurikanien \& Moles, 1989 and the review by Heckman, 1990).

\subsection{HII region-like and LINER nuclei}

The HII nuclei tend to display greater values of $L_N$ and $S_N$ than the
extensive sample of the unclassified nuclei (denoted by UNC in Tables 2 and 3),
which also comprises the few objects classed as L/H and T. On the other hand,
the LINERs have $L_N$ and $S_N$ distributions similar to those of the
unclassified nuclei. In agreement with these results, a straightforward
comparison between the HII and LINER nuclei reveals that the former category
has typically greater values of $L_N$ and $S_N$ than the latter. This holds
also for the two respective subsets of non-interacting objects alone.

Our results suggest that the MIR luminosity of LINERs is indistinguishable from
that of normal (emission-line free) nuclei (which probably dominate the
category of unclassified nuclei), in disagreement with the contention of
Willner et al. (1985), who claimed that LINERs are more luminous than normal
at $\sim$10\m\, on the basis of fewer objects. Furthermore, our results
indicate that the starburst phenomenon leads to an appreciable increase in
MIR emission, although the effect is certainly smaller than that related to
Seyferts.

Unlike some previous contentions (Devereux, 1987), the MIR emission of HII
nuclei does not turn out to be typically more compact that that of LINER and
unclassified objects. (Table 3 shows that the $C$ distributions of these three
categories of objects do not differ significantly.)

Interestingly, a specific comparison between HII nuclei and LINER nuclei
leads to a strong difference in the same sense for radio continuum
luminosities (Giuricin, Mardirossian \& Mezzetti, 1988a; Giuricin et al.,
1988b) and for the sizes of their central radio sources (Giuricin et al.,
1990; Hummel et al., 1990). It also leads to a weak difference in the
opposite sense for the near-infrared luminosities ( in the J, H, K bands)
of their central ($\sim$1 kpc diameter) (Giuricin et al., 1993).
Furthermore, the near-infrared data of a sample of HII and LINER nuclei
compiled by the latter authors show normal near-infrared colours
(which are fully explainable in terms of emission from late-type evolved
stars) for LINER nuclei, but strong redward deviations from the norm of
the colour indices K-L of a subset of HII nuclei hosted in interacting
galaxies. Redder near-infrared colours than normal are usual results of
near-infrared photometric observations of far-infrared (IRAS) bright
galaxies (Moorwood, V\'eron-Cetty \& Glass, 1986, 1987; Carico et al.,
1986, 1988, 1990), of interacting galaxies (Joseph et al., 1984; Cutri
\& McAlary, 1985; Lutz, 1992), and of prominent starburst nuclei
(Lawrence et al., 1985; Devereux, 1989); but there is as yet no consensus
about the interpretation of this (as discussed in Giuricin et al., 1993).
In the soft X-ray band ($\sim$0.5-4 keV) galaxies with HII and LINER nuclei
exhibit emissions of comparable strengths (Giuricin et al., 1991).

\subsection{Early-type and late-type spirals}

Among the non-Seyfert galaxies, early-type spirals (Sb and earlier, denoted by
EARLY in Tables 2 and 3) turn out to be slightly brighter at $\sim$10\m\ than
late-type ones (Sbc and later, denoted by LATE in Tables 2 and 3) by an average
factor of $\sim$1.7 in $L_N$ and $\sim$1.4 in $S_N$; this occurs
notwithstanding the preference of the HII nuclei to be hosted in late-type
spirals, which would lead to the opposite tendency (and despite
the comparable frequency
of occurrence of interacting objects in the early and late morphological
intervals). We have verified that the same tendency holds for the subsets of
unbarred (or barred) galaxies alone (although its statistical significance
becomes quite low because of poor statistics). This small tendency was already
implicit in the data sample by Devereux et al. (1987) and Devereux (1987).
This effect is certainly too large to be ascribed
to a somewhat larger 10\m\ flux contribution of bulge population
stars, if this contribution is taken to be $\sim$10\% of the flux measured in
the standard H band (at $\sim$1.65\m), according to the estimates of Impey,
Wynn-Williams \& Becklin (1986) and the relevant discussion of Devereux et al.
(1987).

Early-type spirals show much more compact emission than late-type ones.
(We have verified that this tendency holds for all bar types.)

Owing to the negligible stellar emission at MIR wavelengths, our new
finding cannot be a trivial result of the prominence of bulges in
early-type spirals, which accounts for a similar difference in the
central near-infrared luminosities of early- and late-type spirals
(see, e.g., Devereux et al., 1987). Our finding is likely to be more
closely linked with the results of the radio surveys of extensive samples
of generic spiral galaxies. These reveal that early-type spirals contain
stronger and more compact central radio sources than late-type objects
(Hummel, 1981a; van der Hulst, Crane \& Keel, 1981). This difference is
not confirmed in the recent study by Hummel et al. (1990), probably
because their sample, being restricted to galaxies having LINER or HII
nuclei, is skewed towards galaxies with HII nuclei (which are stronger
radio sources than the norm and which occur mostly in late-type spirals).
Incidentally, early-type spirals appear to have far-infrared luminosities
similar to those of late-type ones (Isobe \& Feigelson, 1992).

\subsection{Barred and unbarred spirals.}

Comparing the $L_N$ distribution functions of barred and unbarred spirals via
the Kolmogorov-Smirnov test (e.g., Hoel, 1971), in which upper limits on $L_N$
were replaced simply by detections, Devereux (1987) realized that barred
spirals are typically brighter than unbarred ones in the early-type range (Sb
and earlier), whereas no bar effect was observed in the later types. Our
survival analysis of non-Seyfert galaxies (classed as SA and SB) confirms the
existence of this bar effect in the whole morphological sequence. But, trying
several morphological type subdivisions, we have found that this bar effect is
due essentially to the types Sb, Sbc, Sc ( i.e. T=3, 4, 5) at variance with
Devereux' (1987) claims; as a matter of fact, we have verified that in other
type intervals (such as T$<$3, T$<$4, T$<$5, T$>$3, T$>$4, T$>$5) this bar
effect is weaker and sometimes statistically not significant. Including
the transition bar-type SX into the category of barred (SB)
galaxies, we have verified that unbarred spirals are fainter than
SB+SX objects (in the whole morphological sequence and in the Sb,Sbc,Sc
range).
Interestingly, we have verified that this bar effect is no longer significant
if
we exclude the galaxies with HII nuclei from our sample (the two respective
subsamples are denoted by SA (NON-H) and SB (NON-H) in Tables 2 and 3), whilst
it becomes only a little less significant or
almost equally significant if we exclude
the interacting galaxies or the LINERs from our sample. This suggests that the
association of bars with enhanced 10\m\ emission holds mostly for HII nuclei.

The spatial distribution of the MIR emission appears to be more compact in
barred galaxies (SB and SB+SAB) than in unbarred ones (SA) in the whole
morphological range and in the  types Sb,Sbc,Sc; again, as in our previous
discussion on log $L_N$ and log $S_N$, we have verified that the bar effect on
the compactness of the emission is substantially due to the types Sb,Sbc,Sc,
in disagreement with Devereux (1987), who stressed the predominance of
compact emission in the early types.

Analogously, barred spirals tend to have more powerful central radio
sources than unbarred systems (Hummel, 1981a); later, it was specified
that this effect is substantially confined to galaxies with HII
region-like nuclei only (Giuricin et al., 1990; Hummel et al., 1990),
which is consistent with our results. No bar effect was found in the
near-infrared colours J-H and H-K and in the near-infrared luminosities
$L_J$, $L_H$, $L_K$ of the central regions of non-Seyfert spirals
(Giuricin et al., 1993). On the other hand, the studies of IRAS data
have yielded discordant results. On the basis of a far-infrared selected
galaxy sample, Hawarden et al. (1986) found that barred spirals (which
do not host a LINER or Seyfert nucleus) have enhanced far-infrared
luminosities as well as luminosity ratios $L_{25\mu m}/L_{12\mu m}$
and  $L_{25\mu m}/L_{100\mu m}$ with respect
to unbarred spirals. Bothun et al. (1989) found similar
far-infrared luminosities for optically-selected barred and unbarred
spirals. Lastly, analyzing a volume-limited sample of optically-selected
nearby spirals, Isobe \& Feigelson (1992) claimed that barred spirals have,
on average, fainter far-infrared luminosities than unbarred ones.

\subsection{Ringed and unringed spirals.}

We have examined the effects of the ring types of the standard morphology
classification reported in the RC3 catalogue, comparing the $L_N$ and $S_N$
distributions of the S(r) spirals (which possess inner rings), the unringed
S(s) spirals (with S-shaped arms), the S(rs) (transition type) spirals (denoted
respectively by SR, ST, SS in our tables 2 and 3). No significant effect
was found in the non-Seyfert nuclei (Tables 2 and 3 report some results).

Incidentally, Isobe \& Feigelson (1992) claimed that S(r) galaxies have
weaker far-infrared luminosities than average.

\subsection{Relation between radio and mid-infrared emissions}

The results of our detailed comparison between the behaviour of various
categories of objects clearly emphasize that the 10\m\ emission is
particularly closely linked to the radio continuum emission, which in
non-Seyferts is predominantly non-thermal synchrotron emission coming
from supernova remnants mostly. Hence, our study, based on a wide sample
of data, provides a meanigful, strong support for previous similar
claims based on a detailed comparison of the MIR and radio maps of a few,
well-observed galaxies (e. g., Telesco, 1988; Ho et al., 1989), at
resolution of several arcsec (corresponding to several hundreds pc);
however, the MIR-radio correlation seems to become weaker at the smallest
spatial scales ($<$100 pc) currently reachable at MIR wavelengths in very
nearby galaxies (see, e.g., the high resolution MIR maps of Telesco
\& Gezari, 1982, and Pi\~na et al., 1992 for M82 and N253, respectively).
     For non-AGNs the latter fact would favour mechanisms of thermal
emission by small dust grains transiently heated radiatively by newly
formed, hot stars rather than heated by shocks in supernova remnants
(see, e.g., Telesco \& Gezari, 1982, and Ho et al., 1989 for the two
alternative views, respectively). The grain size distribution is thought
to be altered towards the smallward from the norm through mechanisms
of destruction of the largest grains by grain-grain collisions in
interstellar shocks driven by supernova remnants, although observational
evidence for the occurring of this process in Galactic supernova remnants
is far from being ubiquitous (e.g., the review by Dwek \& Arendt, 1992).
Some observational evidence
consistent with a population of transiently heated small grains
(probably depleted in the most intense areas of starbursts) comes
out from the spatial variation of
MIR colour indices across the central regions of M82 (Telesco, Decher \&
Joy, 1989) and N253 (Pi\~na et al., 1992), from the anomalously high
10\m\--to--thermal radio flux ratio (e.g., Wynn-Williams \& Becklin, 1985),
and from the frequent presence of UIR emission bands in the nuclear MIR
spectra of starburst nuclei (Roche et al., 1991). The UIR bands are generally
attributed to emission by small grains or large molecules , like the
polyciclic aromatic hydrocarbon (PAH) molecules (e.g., the review by
Puget \& L\'eger (1989)); the UIR features are often absent in AGNs
(Roche et al., 1991), probably because they are destroyed by the intense
hard (EUV and X-rays) radiation field from the central source (Voit, 1991,
1992).

\subsection{Relation between X-rays and mid-infrared.}

It is known that the 10\m\ emission of galactic nuclei correlates with the
radio continuum emission (e.g., Telesco, 1988), with the near-infrared emission
(e.g., Cutri \& McAlary, 1985; Devereux, 1989; Lutz, 1992), and with the
far-infrared emission (at least for objects believed to be powered by
starbursts; see, e.g., Scoville et al., 1983; Telesco, 1988). In this
subsection we wish to explore whether it correlates also with the X-ray
emission
in Seyferts, whose X-ray emission is generally dominated by a bright
nuclear source. This issue has been very little discussed
in the literature.

We have gathered together the soft X-ray fluxes $F_{sX}$
of the Seyferts observed with the ``Einstein Observatory''
satellite (in common with our sample) from the compilation of
Fabbiano, Kim \& Trinchieri (1992) (and references cited therein). Most of
these X-ray data were obtained with the Image Proportional Counter
instrument and in their catalogue were converted into 0.2-4 keV fluxes.
In the case of multiple entries for a galaxy we have adopted the mean
values of the fluxes $F_{sX}$.

Owing to the presence of upper limits on the MIR data and to
the smallness of the sample, we have used the Cox (C) and Kendall (K)
correlation
tests (mentioned in \S 3) in order to compute the correlation probabilities
between the fluxes $F_N$ and $F_{sX}$ for 14 Seyferts (N1068, N1365,
N1386, N2992, N3227, N3783, N4051, N4151, N4388, N4639, N5506, N6814, N7213,
N7582). We found no significant correlations.
Fig. 6 shows the log $F_N$ -- log $F_{sX}$ plot.
The study of the relation between the central MIR and
the soft X-ray relation may be plagued by the fact that in some
Seyferts the X-ray fluxes probably refer to a region larger
than the area encompassed by the small-beam MIR
observations. Observations with high spatial resolution in X-rays would be
required to test adequately the MIR -- X-ray correlation.
Furthermore, internal obscuration of soft X-rays may be a potential source of
considerable scatter in the correlation for both galaxy samples. But this
is a minor problem, if it is true that X-ray bright AGNs suffer, on average,
less internal obscuration than X-ray faint objects do (see,e.g., Reichert et
al.,1985; Turner \& Pounds,1989), perhaps because high-luminosity nuclei
make a more hostile environment for absorbing matter.

Compared to soft X-rays, hard X-rays suffer little attenuation by interstellar
matter and are good indicators of non-thermal emission because rapid
variability and energy arguments imply that they originate very near the
central engine. Turning to the X-rays of higher energies, we have considered
basically the HEAO-1 A2 X-ray (2-10 keV) flux-limited sample of Piccinotti et
al. (1982) (for N2992, N3783, N4151, N4593, N5506, N7172, N7213) complemented
by the HEAO-1 A2 (2-10 keV) fluxes ($F_{hX}$) (or 2$\sigma$ upper limits)
evaluated by Della Cecca et al. (1990) (for N1068, N3227, N3982, N4051, N5929).
We have taken the average of the data for the first and second scan given by
Piccinotti et al. (1982) in order to get fluxes in good agreement with those
derived by Della Cecca et al. (1990) through a different procedure, except in
the case of N3783 (for which we adopted the given upper limit). To this sample
we have added the 2-10 keV fluxes of N4388 observed with the University of
Birmingham X-ray telescope on the Spacelab-2 mission (Hanson et al., 1990) and
the average of the HEAO-1 A2 (2-20 keV) fluxes of the X-ray variable galaxy
N6814 observed by Tennant et al. (1981) (The recent GINGA observations of N6814
yielded very similar 2-20 keV fluxes (Kunieda et al., 1990)). This sample of 15
Seyferts with hard X-ray data is hereafter referred to as sample A.

We have assembled another sample of hard X-ray data (hereafter referred to as
sample B) substituting the data of Piccinotti et al. (1982) and Della Cecca et
al. (1990) with the EXOSAT 2-10 keV fluxes (corrected for intrinsic and
Galactic absorption) given by Turner \& Pounds (1989). Although short- and
long-term X-ray variability is a common phenomenon in AGNs, the fluxes do not
generally vary by more than a factor 2 (see, e.g., Grandi et al., 1992). Fig. 7
and 8 illustrate the log $F_N$ -- log $F_{hX}$ plot for samples A and B of
15 Seyferts. Owing to the presence of two cases of double censoring (both on
$F_N$ and $F_{hX}$) we have applied only the generalized Kendall tau test
to the log $F_N$ -- log $F_{hX}$ . Again we have detected
no appreciable correlations between the fluxes. Table 4 contains the results
of our correlation analysis ($N_{T}$ is the total number of objects,
$N_{L}$, $N_{L'}$, $N_{L"}$ are the numbers of upper limits on the
independent variable, dependent variable, and both variables, respectively).

For AGNs, several authors have reported a correlation between the central
near-infrared emission and the X-ray emission (see, e.g., the recent
results of Kotilainen et al., 1992 and of Danese et al., 1992) and
between the global far-infrared and X-ray emissions (e.g., Green,
Anderson \& Ward, 1992; David, Jones \& Forman, 1992), although a close
inspection of the paper by Green et al. (1992) reveals that the latter
correlation is substantially induced by radio-loud quasars and disappears
in a subsample of radio-quiet quasars and Seyferts. To our knowledge,
there is only one paper (Carleton et al., 1987) in which the
correlation between ground-based, small-beam 10\m\ measures and X-ray emission
is addressed. For a heterogeneous sample of detected AGNs (quasars and
Seyferts), dominated by the X-ray selected sample of Piccinotti et al. (1982),
Carleton et al. (1987) claimed that the luminosities in the two bands
correlate, in disagreement with our results (probably because his sample
contains also several quasars).

The infrared -- X-ray correlation is often cited as one of the main
arguments supporting a non-thermal origin for most of the infrared
continuum  of AGNs (see, e.g., Bregman, 1990, for relevant debates).
The lack of correlation between the 10\m\ and X-ray emissions of Seyferts
is not in favour of the non-thermal origin of the MIR emission of Seyferts.
(This conclusion may not hold for quasars). Other arguments against
the non-thermal model (specifically for the MIR emission of Seyferts)
are the lack of MIR (as well as far-infrared) variability in radio-quiet
quasars and Seyferts (Edelson \& Malkan, 1987; Neugebauer et al., 1989)
and the sizes of the smallest nuclear MIR sources observed (see Telesco
et al., 1984 for N1068 and Neugebauer et al., 1990 for N4151); these sizes
are easily consistent with the order-of-magnitude predictions
($\sim$10-100 pc) of various models of thermal emission by central dust
heated by the central continuum source (Barvainis, 1987; Sanders et al.,
1989; Laor \&Draine, 1993), whereas synchrotron models of MIR emission
would require sources smaller than 10$^{-4}$pc.

\section{Conclusions}

Our statistical analysis has revealed a number of previously unrecognized
features of the nuclear MIR emission of spiral galaxies. We
summarize the main results of our study as follows:

i) The Seyfert galaxies contain the most powerful nuclear sources of 10\m\
emission, which in $\sim$1/3 of cases provide the bulk of the emission of the
entire galaxy; MIR emission in the outer regions is not uncommon in Seyferts.

ii) Interacting nuclei are brighter at $\sim$10\m\ than non-interacting
ones, although the effect is less pronounced than is generally believed. In
addition, our study suggests that the enhancement in the central MIR emission
may be essentially due to an increase of the central MIR surface brightness in
the case of interacting LINERs and to a spread of the central MIR emission over
a larger area in the case of interacting HII nuclei (similarly to
the interaction-induced enhancement of the central radio emission of
LINER and HII nuclei). The increased compactness of
the 10\m\ emission of interacting objects agrees with the similar
behaviour of several indices of star formation activity.

iii) Among the non-Seyferts, HII region-like nuclei are, on average,
stronger emitters at $\sim$10\m\ than normal nuclei (which confirms
previous results) and than LINER nuclei, whose level of emission
is not distinguishable from that of normal nuclei (this is a new result).

iv) Early-type spirals have stronger and more compact 10\m\ emission than
late-type spirals.

v) Barred spirals have stronger and more compact 10\m\ emission than unbarred
systems, essentially because they more frequently contain HII nuclei.

vi) Ringed and unringed spirals have similar central 10\m\ emission.

vii) The 10\m\ emission of Seyferts appears to be unrelated to their
X-ray emission, at variance with some previous claims.

The results of our detailed comparison between the behaviour of various
types of objects clearly emphasize that the 10\m\ emission is
particularly closely linked to the (predominantly non-thermal synchrotron)
radio emission (at least at spatial scales of several hundreds pc).

\bigskip
\bigskip

The authors thank A. Biviano and M.Girardi for help in the statistical
analysis and the referee J. L. Turner for constructive comments.
The authors are grateful for the ASURV software package
kindly provided by E. D. Feigelson. This work has been partially
supported by the Italian Research Council (CNR-GNA) and by the Ministry
of University, Scientific, and Technological Research (MURST).

\newcommand{\apj}{{\it ApJ}}
\newcommand{\apjs}{{\it ApJS}}
\newcommand{\aj}{{\it AJ}}
\newcommand{\mnras}{{\it MNRAS}}
\newcommand{\aea}{{\it A\&A}}
\newcommand{\aeas}{{\it A\&AS}}
\newcommand{\pasp}{{\it PASP}}
\newcommand{\araea}{{\it ARA\&A}}

\section*{Figure captions}

\bigskip

{\bf Figure 1:} The log $L_N$-log A plot, where the 10\m\ luminosity $L_N$ is
in
solar units and the projected diameter A of the central region is in kpc.
Different symbols denote the Seyferts, the HII region-like nuclei, the LINERs
and the other objects. Upper limits on $L_N$ are denoted by arrows. The solid
line is the mean linear regression line.

\bigskip
{\bf Figure 2:} The C-A plot, where C is the compactness parameter. Symbols
as in Fig. 1. The dashed lines represent a grid of simple galaxy models in
which the surface brightness decreases as $e^{-r/r_s}$ where $r_s$ is the
characteristic radius (in kpc) of the surface brightness distribution of the
galaxy.

\bigskip
{\bf Figure 3:} The log $S_N$-log A plot, where $S_N$ is the 10\m\ surface
brightness (in $L_{\odot}$ kpc$^{-2}$ ). Symbols as in Fig. 1.

\bigskip
{\bf Figure 4:} The C-log $L_N$ plot. Symbols as in Fig. 1.

\bigskip
{\bf Figure 5:} The C-log $S_N$ plot. Symbols as in Fig. 1.

\bigskip
{\bf Figure 6:} The log $F_N$--log $F_sx$ plot, where the 10\m\ fluxes $F_N$
are in units of mJy and the soft X-ray fluxes $F_sx$ are in erg
cm$^{-2}$s$^{-1}$. Upper limits are denoted by arrows.

\bigskip
{\bf Figure 7:} The log $F_N$-log $F_hx$ plot, where the 10\m\ fluxes $F_N$ are
in
units of mJy and the hard X-ray fluxes are in erg cm$^{-2}$s$^{-1}$ ,
for  sample A of Seyferts. Upper limits are denoted by arrows.

\bigskip
{\bf Figure 8:} The log $F_N$-log $F_hx$ plot, where the 10\m\ fluxes are
in units of mJy and the hard X-ray fluxes are in erg cm$^{-2}$s$^{-1}$ ,
for sample B of Seyferts. Upper limits are denoted by arrows.

\end{document}